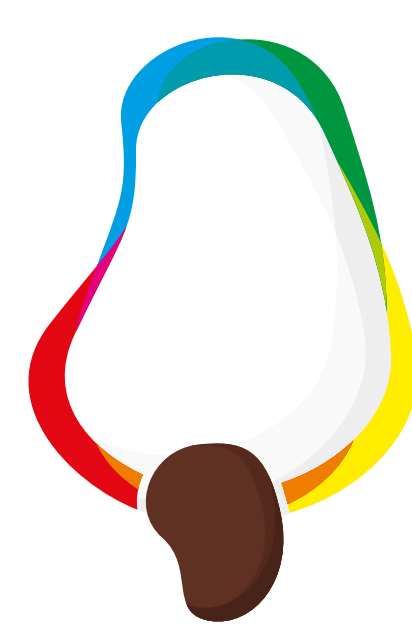

# Germination capacity of pistachio (*Pistacia vera* L.) seeds related to genotypic variation and phytochemical contents

Aram Akram Mohammed[1]*, Ibrahim Maaroof Noori[1]

**Abstract:** Genetic diversity and phytochemical components are the endogenous factors that influence seed germination. The current study aimed to compare the seed germination capacity of 15 *Pistacia vera* genotypes after assessing their genotypic variation using 32 primers (16 ISSR and 16 RAPD) and phytochemical contents. The obtained results explained that the ISSR primers classified the 15 *P. vera* genotypes into four groups, while the RAPD primers classified them into three groups. The genotypes G11, G5, G1, G9, G6, G14, and G10 had the highest germination percentages (98.89, 97.67, 96.67, 94.44, 93.33, 93.33, and 91.11%), respectively. Additionally, their germination speeds were also the highest. However, the lowest germination percentages (62.22 and 68.59%) were recorded in G8 and G4, respectively. Meanwhile, (G9, G10, and G11), (G1 and G14), and (G5 and G6) were identified together in the same group in accordance with both ISSR and RAPD primers. Also, G4 and G8 were in the same subgroup based on RAPD primers. Moreover, the maximum percent protein values (21.88, 21.88, and 20.78%) were measured in the seed kernels of G9, G11, and G1, respectively. Soluble sugar content was the best (798.9 µg $g^{-1}$) in G11. The best percentage of oil (45.3%) was observed in G5.

# Capacidade de germinação de sementes de pistache (*Pistacia vera* L.) relacionada à variação genotípica e aos conteúdos fitoquímicos

**Resumo:** A diversidade genética e os componentes fitoquímicos são os fatores endógenos que influenciam na germinação das sementes. O presente estudo teve como objetivo comparar a capacidade de germinação de sementes de 15 genótipos de *Pistacia vera* após avaliar sua variação genotípica usando 32 primers (16 ISSR e 16 RAPD) e seus conteúdos fitoquímicos. Os resultados obtidos mostram que os primers ISSR classificaram os 15 genótipos de *P. vera* em quatro grupos, enquanto os primers RAPD os classificaram em três grupos. Os genótipos G11, G5, G1, G9, G6, G14 e G10 tiveram as maiores porcentagens de germinação (98.89; 97,67; 96,67; 94,44; 93,33; 93,33 e 91,11%), respectivamente. Além disso, suas velocidades de germinação também foram as mais altas. No entanto, a menor porcentagem de germinação (62,22 e 68,59%) foi registrada para os genótipos G8 e G4, respectivamente. Enquanto isso, os genótipos G9, G10 e G11, G1 e G14 e G5 e G6 foram identificados no mesmo grupo, de acordo com os primers ISSR e RAPD. Além disso, os genótipos G4 e G8 estavam no mesmo subgrupo com base nos primers RAPD. Além disso, os valores máximos de porcentagem de proteína (21,88, 21,88 e 20,78%) foram obtidos nas sementes dos genótipos G9, G11 e G1, respectivamente. O teor de açúcar solúvel foi o mais alto (798,9 $\mu g\ g^{-1}$) no genótipo G11. A maior porcentagem de óleo (45,3%) foi observada no genótipo G5.

[1] Horticulture Department, College of Agricultural Engineering Sciences, University of Sulaimani, Sulaymaniyah city, Kurdistan Region, Iraq.
*Corresponding author: aram.hamarashed@univsul.edu.iq

# Introduction

The seed of *Pistacia vera* L. from Anacardiaceae is the only commercially edible one among 11 species of the *Pistacia* genus. It consists of a kernel inside and surrounded by a hard shell (KASHANINEJAD; TABIL, 2011). Physiological maturity of pistachio seed is determined by shell splitting on the ventral suture and easy separation of the hull which turns blue-green, while seeds with pinkish hulls should be avoided because they are blank (HARTMANN et al., 2011). The seed is mostly germinated for breeding purposes or to produce rootstocks. The dioecy of *P. vera* leads to the impossibility of determining the sex type of the seedlings that germinate from seeds (MOHAMMED; ARKWAZEE, 2024). Besides, seed propagation does not provide true-to-type progenies because of the variabilities that occur in the genetic matter of the seeds during formation, and it delays fruit bearing as well (SHEIKHI et al., 2019). From the germination point of view, seeds of *P. vera* do not have the dormancy phenomenon as other *Pistacia* species, however some external and internal factors may influence the germination parameters and subsequent growth traits of the produced seedlings (BASHABSHEH et al., 2018). Genotype is one of the factors that impact on seed germination. Islam et al. (2009) referred that the important step in crop production is seed germination, which largely relies on genetic factors that control germination parameters and subsequent seedling growth. To reveal the genetic diversity among genotypes, ISSR and RAPD primers have been frequently applied to *P. vera* (JAVANSHAH et al., 2007; NOROOZI et al., 2009; BAGHIZADEH et al., 2010).

Seed germination capacity is varied among genotypes of the same species. This variation arises from differences in physiological and biochemical processes alongside germination that stem from genetic diversity (BRUNEL et al., 2009). The expression and efficiency of key enzymes involved in seed germination, such as α-amylase, proteases, and lipases are reportedly genotype-dependent that collectively provide energy for germination through the breakdown of seed stored food reserves (HELLEMANS et al., 2018). Apart from this, Wilson (2004) stated that seed germination essentially depends on the reserved foods and other phytochemicals stored in the seed, and these stored material ratios are different according to genotype. Hence, seeds of different genotypes may respond differently to the same germination inducer (HE et al., 2014). In this regard, Rabadán et al. (2019) found the effect of genotype on the levels of stored oil and other components in pistachio seeds. Besides, links have been established between seed physical characteristics, particularly seed size and weight, and germination capacity together with the vigor of the produced seedlings (TABAKOVIC et al., 2020). Differences in seed physical characteristics originate from genetic heredity (REZAEI et al., 2019). Maghdouri et al. (2021) indicated that understanding germination is difficult by evaluating one or two factors because it is a complex process; therefore, to obtain an optimum germination percentage, multidimensional strategies should be adopted along with the analysis of diverse factors. In the present study, seed germination capacity and subsequent seedling growth characteristics of different genotypes of *P. vera* were investigated concerning genetic diversity among germinated seeds, seed physical characteristics, seed-reserved food and phytochemical components.

# Materials and Methods

## *Collection of* P. vera *seeds*

Seeds of 15 genotypes of *P. vera* were collected in two different orchards existing in Halabja and Sulaymaniayh governorates (Figure 1) in September 2023 depending on morphological variations, such as shape, color, and size. The split nuts were selected, the hulls were removed, and dried under the sun, then they were coded as G1 to G15 (Figure 2), kept in paper bags and stored at room temperature until the time of sowing.

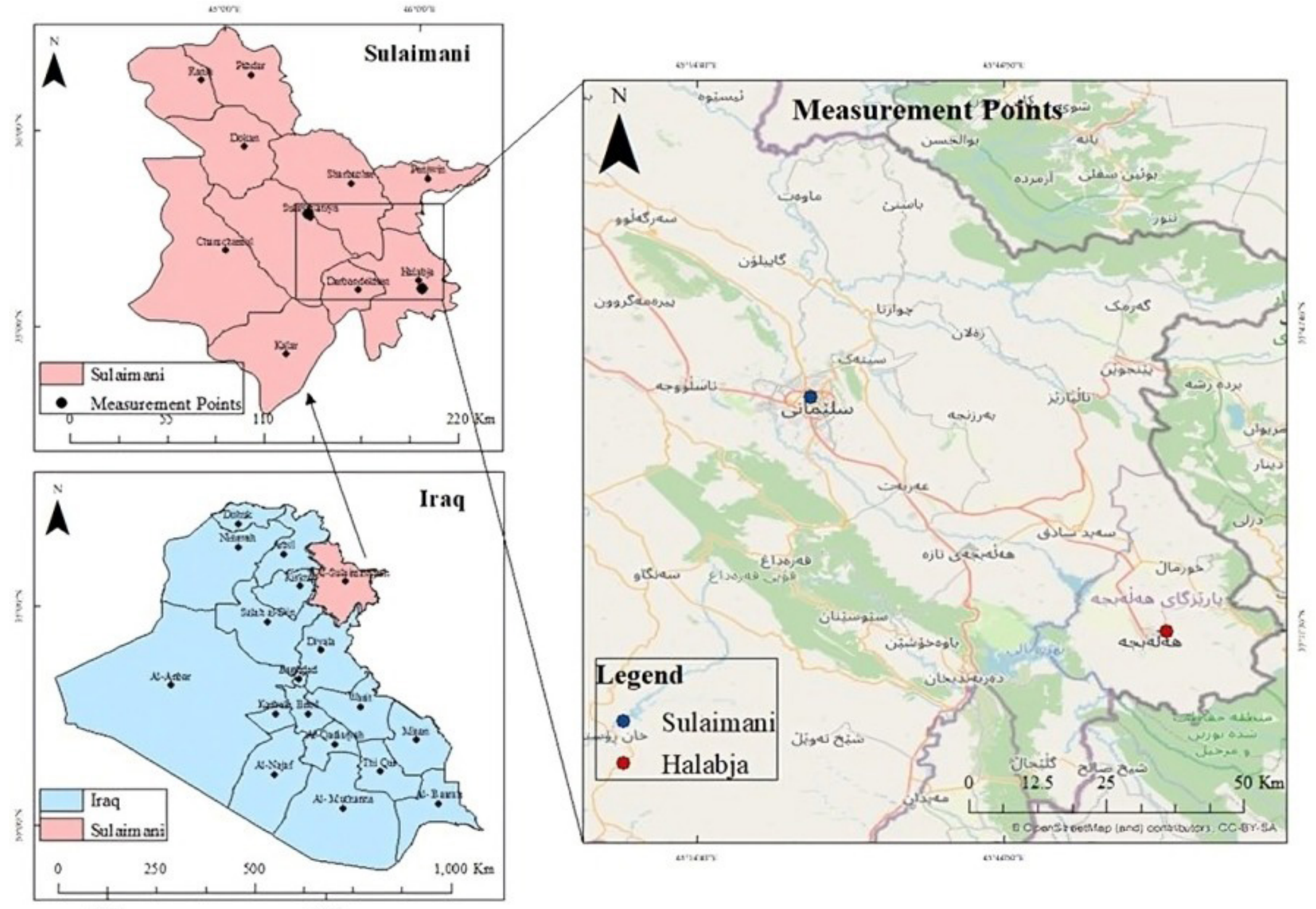

**Figure 1.** Geographic map of the two orchards in Halabja and Sulaymaniayh governorates.

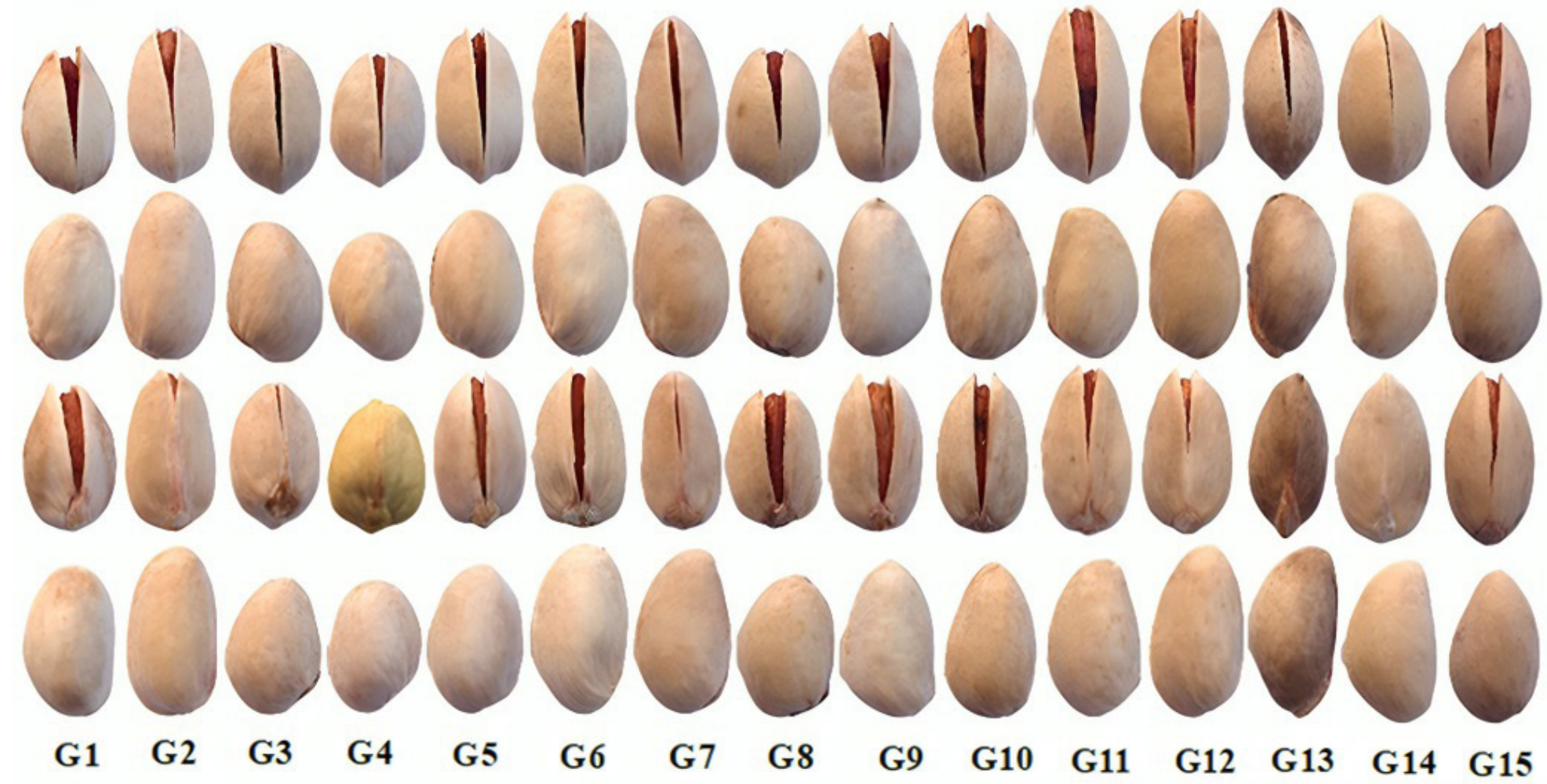

**Figure 2.** Seeds of the 15 *P. vera* genotypes studied, showing some morphological aspects.

## DNA extraction

During late fall (2023), several seeds of each *P. vera* genotype were sown in peat moss media under a greenhouse condition at the College of Agricultural Engineering Sciences, University of Sulaimani. After seed germination and seedling emergence, the fresh leaves were collected, pulverized in liquid nitrogen, then the cetyl trimethyl ammonium bromide (CTAB) protocol was used for the extraction of DNA (KOUAKOU et al., 2022; RASUL et al., 2022). A nanodrop spectrophotometer (Nano PLUS-MAANLAB AB, Sweden) and 1% agarose gel were used to ascertain the quality and quantity of the extracted DNA.

## RAPD and ISSR assays

The primers were employed to analyze the genetic diversity among the 15 pistachio genotypes including sixteen Inter-Simple Sequence Repeats (ISSR) (AHMED et al., 2023) and sixteen Random Amplified Polymorphic DNA (RAPD) (MOLIN et al., 2013; TAHIR, 2014a; TAHIR, 2014b). The reaction mixture of the PCR contained 4.5 µL DNA template, 9 µL PCR master mix, 5 µL primer, and 4 µL deionized water. Amplification was conducted using (Prime Thermal Cycler), initiated by denaturation at 94 °C, took 10 min, continued for 36 cycles of denaturation at 94 °C for 1 min, annealing for 1 min at a specific temperature according to the primer (Table 1), extension for 2 min at 72 °C, and finally for 10 min at 72 °C. The amplified products were run on agarose gels at 1.6% and ethidium bromide was applied for staining.

**Table 1.** The ISSR and RAPD primers used for the amplification of the DNA of *P. vera* genotypes with their sequences and annealing temperatures

| ISSR | Sequence (5' to '3) | Temp. °C | RAPD | Sequence (5' to '3) | Temp. °C |
|---|---|---|---|---|---|
| UBC-811 | GAGAGAGAGAGAGAGAC | 50 | OPY-10 | CAAACGTGGG | 36 |
| UBC-812 | GAGAGAGAGAGAGAGAA | 50 | OPAW-10 | GTTGTTTGCC | 36 |
| UBC-813 | CTCTCTCTCTCTCTCTT | 50 | OPAV-19 | CTCGATCACC | 36 |
| UBC-815 | CTCTCTCTCTCTCTCTG | 50 | OPAT-08 | TCCTCGTGGG | 36 |
| UBC-823 | TCTCTCTCTCTCTCTCC | 50 | OPAR-05 | CATACCTGCC | 36 |
| UBC-825 | ACACACACACACACACT | 50 | OPAR-15 | ACACTCTGCC | 36 |
| UBC-834 | AGAGAGAGAGAGAGAGGT | 50 | OPAM-01 | TCACGTACGG | 36 |
| UBC-841 | GAGAGAGAGAGAGAGACTC | 50 | OPAD-14 | GAACGAGGGT | 36 |
| UBC-845 | CTCTCTCTCTCTCTCTTG | 50 | OPAX- 10 | CCAGGCTGAC | 36 |
| UBC-846 | CACACACACACACACAAT | 50 | OPD-02 | GGACCCAACC | 37 |
| UBC-849 | GTGTGTGTGTGTGTGTGA | 50 | OPD-10 | GGTCTACACC | 37 |
| ISSR7 | AGATCCTCCTCCTCCTCC | 50 | OPD-18 | GAGAGCCAAC | 37 |
| ISSR8 | ATCACACACACACACACA | 50 | OPD-20 | ACCCGGTCAC | 37 |
| ISSR9 | CACACACACACACACATG | 50 | OPG-10 | AGGGCCGTCT | 37 |
| ISSR10 | GAGAGAGAGAGAGAGAGG | 50 | OPG-12 | CAGCTCACGA | 37 |
| ISSR14 | GAGAGAGAGAGAGAGAGC | 50 | OPG-14 | GGATGAGACC | 37 |

## Seed morphological measurements

Morphological measurements of *P. vera* seeds from the 15 genotypes were taken by selecting 15 split seeds of each genotype divided into three lots containing 5 seeds, and every lot was considered as a replicate. Seed length, seed height, and seed width were measured from farthest point to point (Figure 3). The seeds were submerged in water inside a graded cylinder and the vol-ume of the seeds was determined by calculating the volume of the raised water. The nut shells were transversely cut in the middle into two halves and used to measure shell thickness at three different positions of both halves along cut faces. An electronic digital caliper was used to take seed dimension measurements. The means were compared using Duncan's Multiple Range Test (P≤0.05) in a Randomized Complete Block Design (RCBD) layout.

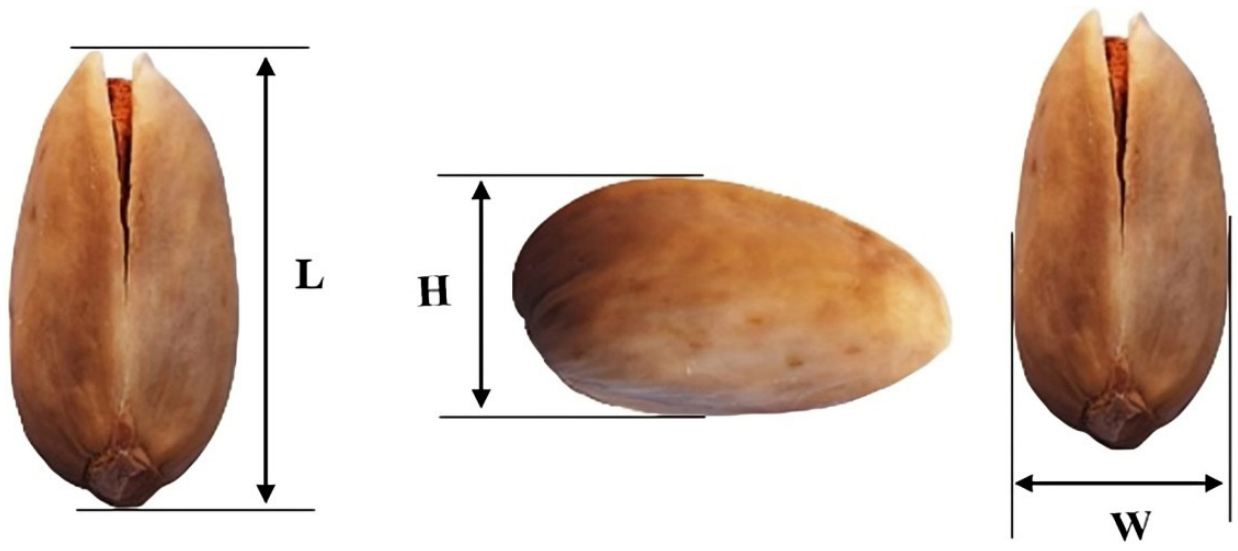

**Figure 3.** Dimensions of *P. vera* seeds. L: seed length, H: seed height, and W: seed width.

## Protein content in seeds

The seed kernels of *P. vera* genotypes were ground in a blender and used to quantify total proteins according to the Johan Kjeldahl method described by Goyal et al. (2022).

## Total carbohydrates in seeds

Total carbohydrate contents of seed kernels of *P. vera* genotypes were quantified according to phenol-sulfuric acid method elaborated by Albalasmeh et al. (2013).

## Oil percentage in seeds

The oil was extracted from the ground kernel of the studied *P. vera* genotypes using extraction technique in *n*-hexane as solvent then the oil percentage was found according to Yahyavi et al. (2020).

## Seed germination and subsequent seedling growth characteristics

Seed germination and subsequent seedling growth characteristics in the 15 *P. vera* genotypes were estimated by selecting 60 healthy split seeds from each genotype. After sterilization by thiophanate methyl fungicide (1 g.L$^{-1}$), they were sown on March 2, 2024 in seed trays (32 cells, with cell diameters of 4×5.5 cm and cell depth of 7 cm) containing peat moss media (Agaris Professional, green color package) in three replications with 20 seeds per replicate, and the experiment was laid out in RCBD inside a greenhouse. The daily minimum and maximum temperatures inside the greenhouse are shown in (Figure 4). The seeds were daily checked for recording germination speed (GAIROLA et al., 2011) and percentage. The parameters of subsequent seedling growth; main root number, taproot length, shoot length, shoot diameter, leaf number, leaf area, total phenols, total soluble sugars, and chlorophylls *a* and *b* (SU et al., 2010) were measured in the seedling after 40 days from the sowing date, in which three seedlings were examined per replicate.

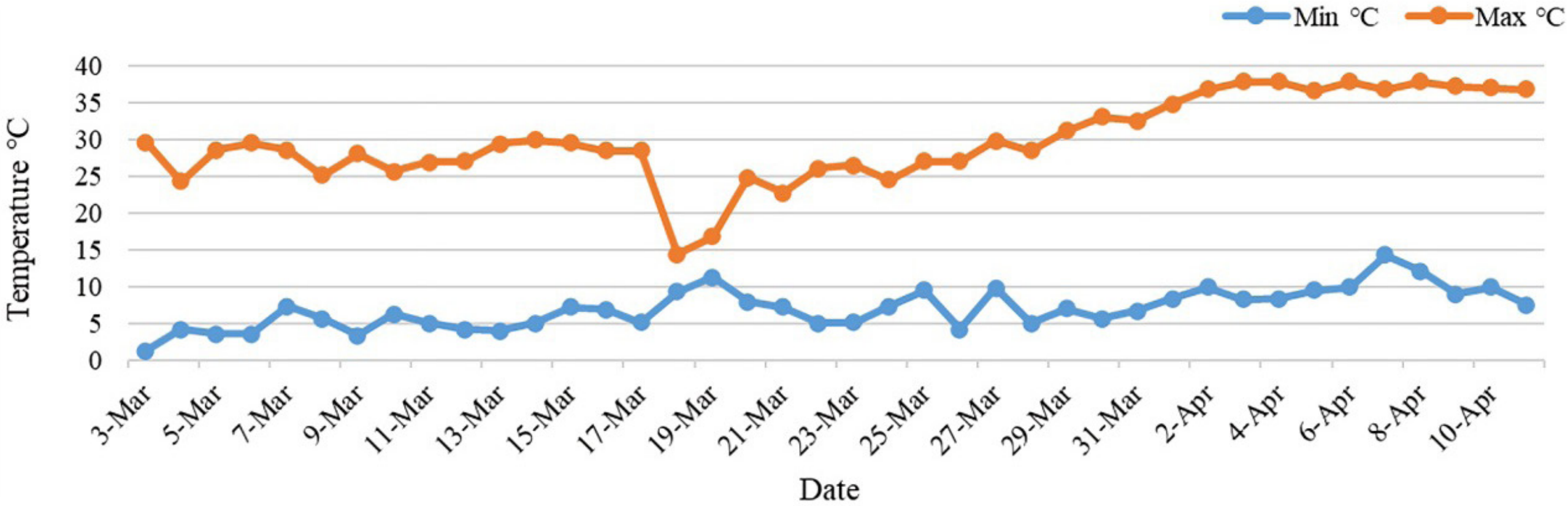

**Figure 4.** Daily minimum and maximum temperatures inside the greenhouse during the germination process from March 3 to April 11, 2024.

### *Contents of total phenols and antioxidants*

Total phenols were measured in both shoots of germinated seeds and seed kernels by weighing 0.1 g of pulverized shoots from germinated seeds in liquid nitrogen, and 0.1 g of the ground kernel in a blender and put in Eppendorf tubes with a capacity of 2 mL. Then, 1 mL methanol 80% was added, shaken for 30 min, and left overnight in the refrigerator. After that, they were centrifuged for 5 min at 10000 rpm, and then the supernatants were collected. From the collected supernatants, 5 µL of the shoot samples and 30 µL of seed samples were taken and mixed with 1.05 mL of Folin–Ciocâlteu reagent at 1:9 and left for 5 min, later 850 µL of $Na_2CO_3$ was added. Finally, the samples were read at 750 nm using a spectrophotometer (UVM6100, MAANLAB AB, Sweden). On the other hand, antioxidant was determined only in seed samples initiated by taking 50 µL of the seed sample supernatants mixed with 2 mL of DPPH prepared at 0.0037 g in 100 $mL^{-1}$ of 95% methanol. The absorbance was read at 517 nm using a spectrophotometer (UVM6100, MAANLAB AB, Sweden).

### *Total soluble sugars*

The anthrone-sulfuric acid colorimetric method was utilized to analyze total soluble sugars in shoots of germinated seeds and seed kernels of the 15 *P. vera* genotypes, in which 0.1 g of shoots of germinated seeds and seed kernels were weighed and placed in 2 mL Eppendorf tubes, then 1 mL of $dH_2O$ was added and left in a water bath at 92 °C for 30 min. After cooling, the samples were centrifuged for 7 min at 7000 rpm, however, the seed kernel samples were double centrifuged in order to further precipitation. Finally, 5 µL from the supernatant of the seed kernel and 15 µL of the supernatant of the shoot samples were taken and mixed with anthrone at 0.225 g in 150 mL of 84% $H_2SO_4$, and once again left in the water bath for 5 min. The absorbance was read at 620 nm using a spectrophotometer (UVM6100, MAANLAB AB, Sweden).

### *Statistical analysis*

Molecular analysis was initially scored manually; 0 for absent and 1 for present amplified bands. Genetic diversity was calculated by PowerMarker software version 3.25. Genetic distances among the genotypes were displayed by Unweighted Pair Group Method Analysis (UPGMA). Seed morphological traits, seed phytochemical components, germination parameters, and characteristics of subsequent seedling growth were analyzed in the RCBD with three replications. XLSTAT version 2020.1.3 was used to UPGMA analyses, and means comparison according to Duncan's Multiple Range Test, Principal Component Analysis (PCA), and Pearson correlation test (P≤0.05). However, Ward method was applied to hierarchical clustering analysis using JMP Pro 16.

## Results

### *Using ISSR and RAPD primers for assessment of genetic diversity*

Genetic diversity was evaluated among the 15 genotypes of *P. vera* (Table 2). The mean values of the number of observed alleles (Na), the effective number of alleles (Ne), expected heterozygosity or gene diversity (He), Shannon's information index (I), and unbiased expected heterozygosity (uHe) were 1.30, 1.35, 0.30, 0.20, and 0.24, respectively, when ISSR primers were used. The highest value of Na (1.63) was recorded as a result of amplification due to UBC 834 primer, at the same time UBC 849 had the lowest Na (1.00). Besides, the values of Ne (1.45 and 1.46), I (0.41 and 0.38), He (0.27 and 0.26), and uHe (0.33 and 0.31) were the maximum as UBC 834 and ISSR 9 primers were respectively used. The primer UBC 849 showed the minimum values of Ne (1.21), I (0.16), He (0.11), and uHe (0.12). The same table shows that Na, Ne,

I, He, and uHe had mean values of 1.51, 1.44, 0.37, 0.25, and 0.28, respectively, under RAPD primers. OPD-18, OPAV-19, and OPAW-10 primers resulted in the best Na (1.83), Ne (1.61), I (0.48), and uHe (0.38), respectively, and also the three RAPD primers had the same high values of He (0.32). The minimum Na (0.89), Ne (1.12), I (0.16), He (0.09), and uHe (0.10) were observed with OPG-14 primer.

**Table 2.** Number of observed alleles (Na), effective number of alleles (Ne), expected heterozygosity or gene diversity (He), Shannon's information index (I), and unbiased expected heterozygosity (uHe) in *P. vera* genotypes using ISSR and RAPD markers

| ISSR markers | Na | Ne | I | He | uHe | RAPD markers | Na | Ne | I | He | uHe |
|---|---|---|---|---|---|---|---|---|---|---|---|
| ISSR 7 | 1.17 | 1.25 | 0.23 | 0.15 | 0.18 | OPD-02 | 1.50 | 1.38 | 0.35 | 0.23 | 0.27 |
| ISSR 8 | 1.08 | 1.33 | 0.28 | 0.19 | 0.23 | OPD-10 | 1.33 | 1.51 | 0.41 | 0.28 | 0.31 |
| ISSR 9 | 1.50 | 1.46 | 0.38 | 0.26 | 0.31 | OPD-18 | 1.83 | 1.56 | 0.47 | 0.32 | 0.36 |
| ISSR 10 | 1.25 | 1.33 | 0.27 | 0.19 | 0.21 | OPD-20 | 1.33 | 1.37 | 0.31 | 0.21 | 0.23 |
| ISSR 14 | 1.25 | 1.33 | 0.28 | 0.19 | 0.23 | OPG-12 | 1.67 | 1.56 | 0.44 | 0.30 | 0.34 |
| UBC 812 | 1.29 | 1.35 | 0.29 | 0.20 | 0.24 | OPG-14 | 0.89 | 1.12 | 0.16 | 0.09 | 0.10 |
| UBC 813 | 1.13 | 1.31 | 0.26 | 0.18 | 0.21 | OPG-10 | 1.67 | 1.50 | 0.42 | 0.28 | 0.32 |
| UBC 814 | 1.21 | 1.35 | 0.28 | 0.19 | 0.22 | OPY-10 | 1.21 | 1.20 | 0.22 | 0.14 | 0.15 |
| UBC 815 | 1.23 | 1.29 | 0.26 | 0.17 | 0.20 | OPAW-10 | 1.80 | 1.54 | 0.48 | 0.32 | 0.38 |
| UBC 823 | 1.50 | 1.38 | 0.34 | 0.23 | 0.27 | OPAV-19 | 1.67 | 1.61 | 0.45 | 0.32 | 0.34 |
| UBC 825 | 1.40 | 1.37 | 0.30 | 0.21 | 0.24 | OPAT-08 | 1.72 | 1.39 | 0.38 | 0.24 | 0.28 |
| UBC 834 | 1.63 | 1.45 | 0.41 | 0.27 | 0.33 | OPAR-05 | 1.67 | 1.48 | 0.41 | 0.28 | 0.31 |
| UBC 841 | 1.39 | 1.38 | 0.34 | 0.23 | 0.27 | OPAR-15 | 1.43 | 1.46 | 0.40 | 0.27 | 0.30 |
| UBC 845 | 1.50 | 1.43 | 0.38 | 0.25 | 0.31 | OPAM-01 | 1.44 | 1.50 | 0.40 | 0.28 | 0.30 |
| UBC 846 | 1.22 | 1.34 | 0.28 | 0.19 | 0.22 | OPAD-14 | 1.43 | 1.40 | 0.35 | 0.23 | 0.26 |
| UBC 849 | 1.00 | 1.21 | 0.16 | 0.11 | 0.12 | OPAX- 10 | 1.58 | 1.37 | 0.35 | 0.23 | 0.25 |
| **Mean** | **1.30** | **1.35** | **0.30** | **0.20** | **0.24** | **Mean** | **1.51** | **1.44** | **0.37** | **0.25** | **0.28** |

The clustering of the 15 genotypes of pistachio into dendrograms (Figure 5) disclosed that there were four different groups based on ISSR primers (Figure 5A). Group 1 (red) included 2 genotypes (G1 and G13). The second group (green) embraced 2 genotypes (G8 and G14). Whereas, Group 3 (blue) contained 6 genotypes and was further divided into two subgroups, the first subgroup having G2 and G3, while G4, G5, G6, and G7 were incorporated into the second subgroup. The genotypes that belonged to the fourth group were 5 genotypes divided into two subgroups; G9, G10, G11, and G15 categorized into the first subgroup, and the second subgroup contained only G12. Additionally, using RAPD primers, the 15 pistachio genotypes classified into three groups (Figure 5B). The UPGMA analysis revealed that Group 1 (red) comprised 6 genotypes with two subgroups; the first subgroup included G1, G12, G13, and G14, while the second subgroup assembled G11 and G15. Group 2 (green) had 7 genotypes with two subgroups. The first subgroup consisted of G2, G4, G5, G6, and G8, meanwhile G3 and G7 were involved in the second subgroup. Only G9 and G10 were included in Group 3 (blue). Looking at Figure (5C) clarifies that the combination of ISSR and RAPD primers distributed the 15 *P. vera* genotypes into three main groups. The number of genotypes in the first group (red) was 4 genotypes divided into two subgroups; G1, G13, and G14 were put in a subgroup and G8 merely in another subgroup. The major group was the second group (green), which had nine genotypes displayed into two subgroups. The first subgroup included G2, G4, G5, G6, G7, and G3, whereas the second subgroup contained G11, G15, and G12. However, G9 and G10 consisted the third group (blue).

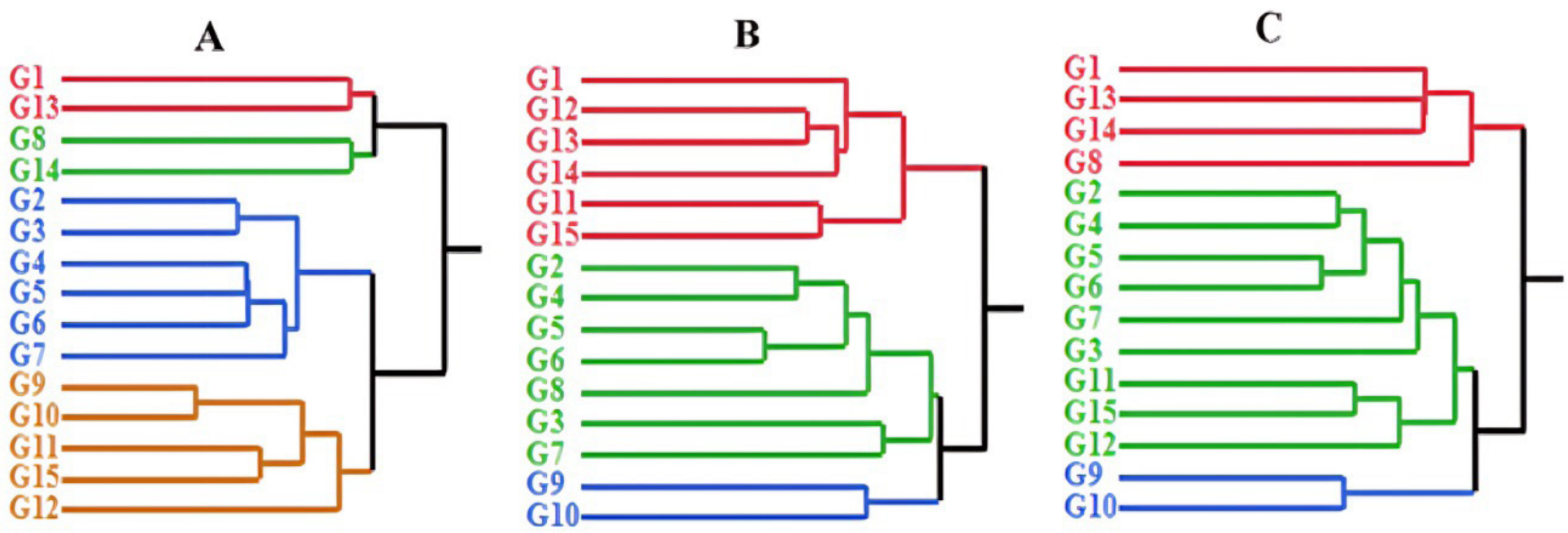

**Figure 5.** Dendrograms of the 15 *P. vera* genotypes created by the Unweighted Pair Group Method with arithmetic mean (UPGMA) according to ISSR (A), RAPD (B), and combination of ISSR and RAPD (C). Each color of the branches represents a group.

The structure analysis of the 15 genotypes of *P. vera* was conducted to determine the number groups of the *P. vera* genotypes and their genetic purities, which were indicated by the value of K against delta K (Figure 6). The highest K = 4 (Figure 6A) signified that *P. vera* genotypes had four groups as a consequence of ISSR primers. The value between 0.20-0.80 verified that four genotypes with red color (G14, G1, G8, and G13) were the first group (Figure 6B). A single genotype with a green color (G12) was marked as the second group. G2, G3, G4, and G6 with blue color were appeared in the third group. The fourth group contained G9 and G10 with yellow color. Meanwhile, G5, G7, G11, and G15 with different colors were considered as admixed and impure genotypes. DNA amplification by RAPD primers identified the 15 *P. vera* genotypes into 3 groups based on the highest value of K at number 3 (Figure 6C). Therefore, the *P. vera* genotypes took three colors (Figure 6D), in which G1, G11, G12, G14, and G13 with red color located in one group. However, green colored genotypes (G4, G5, G6, G7, and G8) were displayed in another group. The last group was composed of G9 and G10 with blue color. Meantime, G2, G3, and G15 were appeared as admixed genotypes. The peak value of K was found at number 3 (Figure 6E) when ISSR primers were combined with RAPD primers. This implies that the *P. vera* genotypes were distributed in three groups (Figure 6F). The first group consisted of G1, G11, G12, G13, and G14 (red). The second, green-colored group had G2, G4, G6, and G7. The third group was composed of G9 and G10 (blue). Whereas, the remaining genotypes (G3, G5, G8, and G15) were shown to be admixed genotypes.

Table (3) explains the genetic diversity indices (Na, Ne, I, He, and uHe) depending on the produced populations (clusters). Using ISSR primers caused the best values of Na (160) in Pop 4 and (159) in Pop 3. Similarly, Pop 3 and Pop 4 had the same highest Ne (149), I (0.41), He (0.28), and uHe (0.30) in Pop 3 and (0.31) in Pop 4. In contrast, Pop 2 brought about the lowest Na (0.90), Ne (1.17), I (0.15), He (0.10), and uHe (0.14). Besides, the primers of RAPD led to the highest Na (1.88), Ne (1.53), I (0.47), He (0.31), and uHe (0.33) in Pop 2. Meanwhile, in Pop 3, Na, Ne, I, He, and uHe were the lowest (1.03, 1.27, 0.23, 0.16, and 0.21, respectively). On the other hand, ISSR markers produced the highest similarity (89%) between cluster 4 (CL4) and cluster 3 (CL3), followed by 83% between cluster 2 (CL2) and cluster 1 (CL1) according to the Nei index (Nei I) (Table 4). The same table also substantiated that CL2 and CL1 were similar by 92% using RAPD markers, and to a lesser extent CL3 and CL2 were similar by 86%.

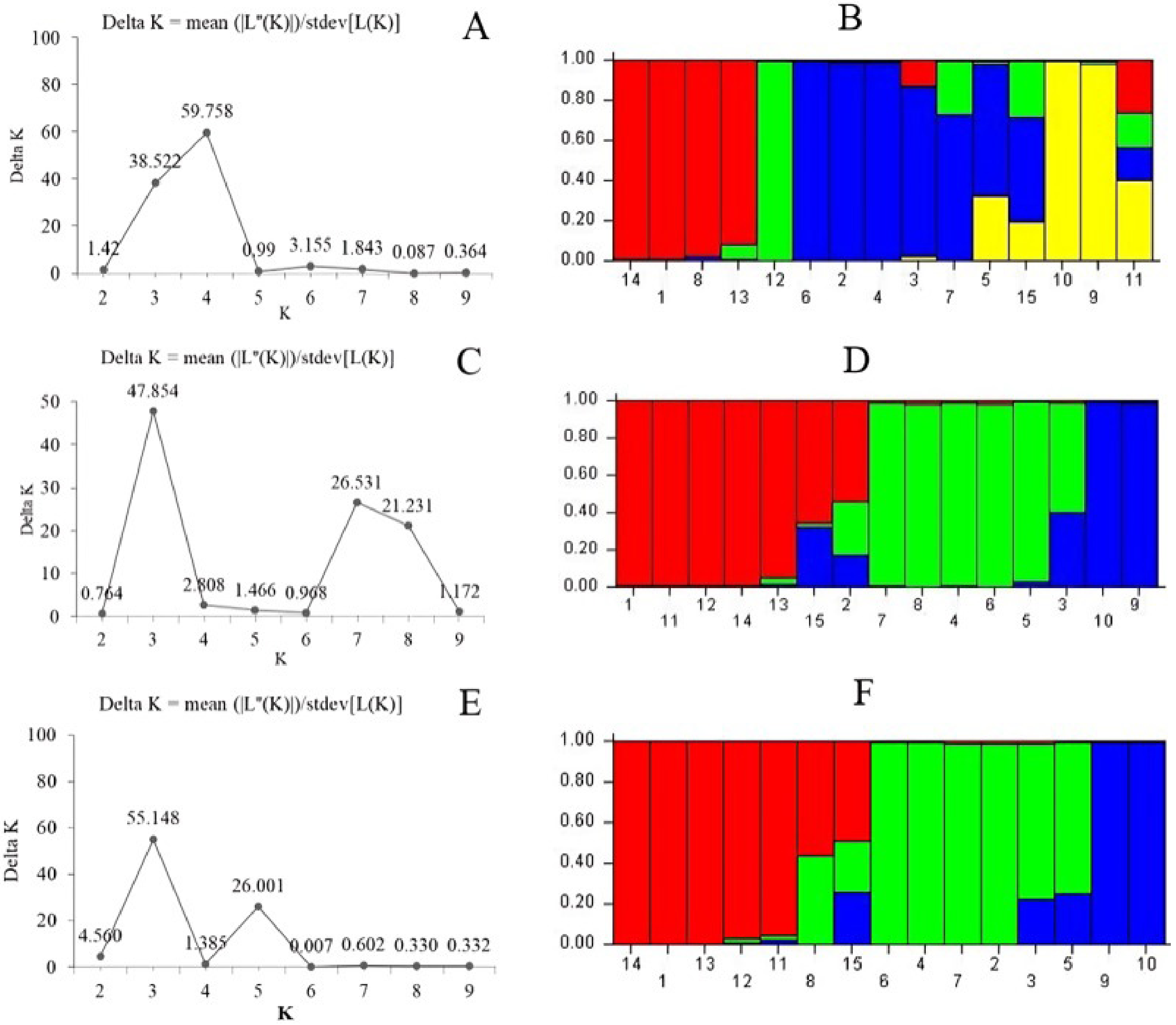

**Figure 6.** Delta K for different group numbers demonstrated by the highest (K) as a result of ISSR (A), RAPD (C), and combination of ISSR and RAPD (E) data. Structure of *P. vera* genotype groups K= 4 (B) from ISSR, K= 3 (D) from RAPD, and K = 3 (F) from ISSR and RAPD data. Each color in (B, D, and F) is representative of a group.

**Table 3.** Determination of diversity indices from ISSR and RAPD markers

| | Pop | Na | Ne | I | He | uHe |
|---|---|---|---|---|---|---|
| **ISSR** | Pop-1 | 1.15 | 1.28 | 0.24 | 0.16 | 0.22 |
| | Pop-2 | 0.90 | 1.17 | 0.15 | 0.10 | 0.14 |
| | Pop-3 | 1.59 | 1.49 | 0.41 | 0.28 | 0.30 |
| | Pop-4 | 1.60 | 1.49 | 0.41 | 0.28 | 0.31 |
| | **Mean** | **1.31** | **1.36** | **0.30** | **0.20** | **0.24** |
| **RAPD** | Pop-1 | 1.69 | 1.48 | 0.42 | 0.28 | 0.31 |
| | Pop-2 | 1.88 | 1.53 | 0.47 | 0.31 | 0.33 |
| | Pop-3 | 1.03 | 1.27 | 0.23 | 0.16 | 0.21 |
| | **Mean** | **1.53** | **1.43** | **0.37** | **0.25** | **0.28** |

(Pop) population, (Na) number of observed alleles, (Ne) effective number of alleles, (I) Shannon's information index, (He) expected heterozygosity or gene diversity, (uHe) unbiased expected heterozygosity.

**Table 4.** Genetic identity between *P. vera* genotype clustering was revealed by the Nei matrix using ISSR and RAPD markers

| | | CL1 | CL2 | CL3 | CL4 |
|---|---|---|---|---|---|
| **ISSR** | **CL1** | 1.00 | | | |
| | **CL2** | 0.83 | 1.00 | | |
| | **CL3** | 0.79 | 0.81 | 1.00 | |
| | **CL4** | 0.78 | 0.73 | 0.89 | 1.00 |

| | | CL1 | CL2 | CL3 |
|---|---|---|---|---|
| **RAPD** | **CL1** | 1.00 | | |
| | **CL2** | 0.92 | 1.00 | |
| | **CL3** | 0.82 | 0.86 | 1.00 |

(CL) cluster.

## *Morphological characteristics and phytochemical components of pistachio seeds*

All seed morphological traits were significantly different in the 15 *P. vera* genotypes investigated in this study. As it is clear in Table (5), G6, G2, G12, and G7 were tops among the genotypes in seed length (22.40, 22.03, 22.00, and 21.72 mm, respectively), whereas the shortest seeds were noticed in G4 and G8 with 17.52 and 17.59 mm, respectively. The highest values of seed height (13.8 mm), shell weight (0.71 g), and shell thickness (0.89 mm) were given by G3. Seed width was the maximum (12.29 mm) in G4. Moreover, G6 had the largest seed volume (2.02 $cm^3$), the greatest seed weight (1.40 g), and kernel weight (0.72 g). Contrarily, the lowest seed height (10.21 mm), seed width (9.77 mm), seed volume (0.98 $cm^3$), and kernel weight (0.40 g) were found in G13. Additionally, G15 and G13 gave the minimum seed weights (0.90 and 0.93 g, respectively). Furthermore, G10 and G9 had the lowest shell weights (0.41 and 0.44 g, respectively) and shell thickness (0.65 mm).

**Table 5.** Seed morphological characteristics of *P. vera* genotypes

| Genotypes | Seed length (mm) | Seed height (mm) | Seed width (mm) | Seed volume ($cm^3$) | Seed weight (g) | Shell weight (g) | Kernel weight (g) | Shell thickness (mm) |
|---|---|---|---|---|---|---|---|---|
| G1 | 18.86 de | 11.15 fgh | 10.55 d | 1.32 de | 1.00 cde | 0.48 de | 0.52 ef | 0.78 cde |
| G2 | 22.03 a | 12.61 bc | 11.45 bc | 1.99 a | 1.33 a | 0.64 b | 0.69 ab | 0.80 bcd |
| G3 | 20.36 b | 13.80 a | 12.00 ab | 1.99 a | 1.33 a | 0.71 a | 0.62 bcd | 0.89 a |
| G4 | 17.52 f | 11.97 c-f | 12.29 a | 1.71 bc | 1.05 bcd | 0.48 de | 0.57 cde | 0.69 f |
| G5 | 19.32 b-e | 11.77 c-g | 10.19 de | 1.49 cd | 0.92 e | 0.48 de | 0.44 gh | 0.81 bc |
| G6 | 22.40 a | 13.00 b | 11.94 ab | 2.02 a | 1.40 a | 0.68 ab | 0.72 a | 0.79 bcd |
| G7 | 21.72 a | 12.27 b-e | 9.98 de | 1.79 ab | 1.13 b | 0.58 c | 0.56 de | 0.77 de |
| G8 | 17.59 f | 11.64 d-h | 11.13 c | 1.63 bc | 0.98 cde | 0.46 de | 0.51 efg | 0.75 e |
| G9 | 19.87 bcd | 10.99 ghi | 10.35 d | 1.53 bcd | 0.96 de | 0.44 e | 0.46 fgh | 0.65 g |
| G10 | 19.33 b-e | 11.01 ghi | 10.48 d | 1.63 bc | 1.01 cde | 0.41 e | 0.59 cde | 0.65 g |
| G11 | 19.97 bc | 12.47 bcd | 10.42 d | 1.44 cd | 1.09 bc | 0.53 cd | 0.56 de | 0.75 e |
| G12 | 22.00 a | 12.20 b-e | 11.65 bc | 1.81 ab | 1.30 a | 0.65 ab | 0.65 abc | 0.82 b |
| G13 | 19.18 cde | 10.21 i | 9.77 e | 0.98 f | 0.93 e | 0.53 cd | 0.40 h | 0.88 a |
| G14 | 19.79 b-e | 10.86 hi | 10.16 de | 1.15 ef | 0.97 de | 0.52 cd | 0.45 fgh | 0.77 de |
| G15 | 18.76 e | 11.43 e-h | 10.05 de | 1.09 ef | 0.90 e | 0.44 e | 0.46 fgh | 0.71 f |

The means in the same column followed by the same letter(s) are significantly not different according to Duncan's Multiple Range Test ($P \leq 0.05$).

Based on the data presented in Table (6), the phytochemicals that were analyzed in the seed kernels of the 15 *P. vera* genotypes were significantly variable. Accordingly, the percentage of carbohydrates got the peaks in G6 (31.78%), G10 (30.80%), and G3 (30.54%), but the lowest carbohydrates (18.72, 18.92, 18.96 and 19.12%) were shown in G5, G12, G1, and G9, respectively. Meanwhile, the highest values of protein percentage (21.88%) were quantified in G3, G9, G11, and G15 accompanied by (20.78%) in G1 and G6. Whereas, proteins were minimal in G5 (17.10%), G13 (17.30%), and both G4 and G8 (17.50%). Furthermore, G5 contained the maximum oil (45.30%) followed by G13 (43.50%) and G1 (43.00%). Concurrently, G3, G10, and G11 incorporated the minimum oil (36.00, 37.00, and 38.00%, respectively). In the same table, it is clear that G11 and to a lesser extent G3 had the best values of soluble sugar contents (798.90 and 628.15 µg $g^{-1}$, respectively). In contrast, the lowest soluble sugar contents (372.40 and 389.10 µg $g^{-1}$) were found in G14 and G9, respectively. The results of total phenol contents clarified that G14 was at the top of the genotypes by 423.37 µg g$^{-1}$ accompanied by G13 (399.78 µg g$^{-1}$) and G5 (373.00 µg g$^{-1}$). Whereas, G6 and G11 showed the lowest total phenols (178.99 and 182.17 µg g$^{-1}$, respectively). Moreover, the best antioxidant values were observed in G14 (1480.81 µg g$^{-1}$), G5 (1478.80 µg g$^{-1}$), and G13 (1469.32 µg g$^{-1}$). Contrarily, G12 and G15 contained the minimum antioxidants (544.32 and 571.00 µg g$^{-1}$, respectively).

**Table 6.** Some phytochemical components in seeds of *P. vera* genotypes

| Genotypes | Carbohydrates % | Proteins % | Oil % | Soluble sugars (µg g$^{-1}$) | Total phenols (µg g$^{-1}$) | Antioxidants (µg g$^{-1}$) |
|---|---|---|---|---|---|---|
| G1 | 18.96 e | 20.78 ab | 43.00 bc | 473.83 fg | 301.27 e | 1359.19 c |
| G2 | 24.80 c | 19.96 bc | 42.00 de | 544.20 de | 251.09 g | 936.22 e |
| G3 | 30.54 a | 21.88 a | 36.00 k | 628.15 b | 234.23 ij | 945.68 e |
| G4 | 28.70 b | 17.50 de | 40.40 gh | 560.25 cd | 301.46 e | 823.38 g |
| G5 | 18.72 e | 17.10 e | 45.30 a | 456.34 g | 373.00 c | 1478.80 a |
| G6 | 31.78 a | 20.78 ab | 40.40 gh | 479.80 f | 178.99 k | 613.24 i |
| G7 | 20.55 de | 19.00 bcd | 41.00 fg | 423.00 h | 318.13 d | 1427.43 b |
| G8 | 26.32 c | 17.50 de | 41.80 def | 576.10 c | 270.75 f | 872.03 f |
| G9 | 19.12 e | 21.88 a | 41.40 ef | 389.10 i | 262.7 fg | 865.95 f |
| G10 | 30.80 a | 19.69 bc | 37.00 j | 435.60 h | 224.12 j | 1001.76 d |
| G11 | 25.53 c | 21.88 a | 38.00 i | 798.90 a | 182.17 k | 753.11 h |
| G12 | 18.92 e | 19.69 bc | 42.40 cd | 528.15 e | 244.91 hi | 544.32 k |
| G13 | 21.17 d | 17.30 de | 43.50 b | 460.04 g | 399.78 b | 1469.32 a |
| G14 | 22.17 d | 18.60 cde | 40.40 gh | 372.40 i | 423.37 a | 1480.81 a |
| G15 | 21.53 d | 21.88 a | 40.00 h | 465.60 fg | 237.60 hij | 571.35 j |

The means in the same column followed by the same letter(s) are significantly not different according to Duncan's Multiple Range Test (P≤0.05).

## Germination parameters and subsequent seedling growth characteristics

Analysis of germination and subsequent growth traits of the 15 *P. vera* genotypes (Table 7) established that the genotypes showed significantly different results of all measured data. It is clear that G11, G5, G1, G9, G6, G14, G10, G15, and G7 had the highest germination percentages (98.89, 97.78, 96.67, 94.44, 93.33, 93.33, 91.11, 86.67, and 85.56%), respectively. The fastest germination was observed in G5, G11, and G10 (11.51, 11.38, and 10.6 seed/time interval), respectively. Also, the longest shoots (13.92, 12.54, 12.19, and 11.33 cm) were measured in G11, G5, G6, and G15, respectively. Moreover, G2 in shoot diameter (2.11 mm) and G10 in leaf number (12.78) were the best genotypes. G6 exhibited the maximum leaf area (8.70 cm$^2$), root number (70.04), and taproot length (14.83 cm). Whereas, the minimum values of germination percentages (62.22, 68.89, and 70%) were recorded in G8, G4, and G12. Besides, germination speed (2.46 and 3.82 seed/time interval), shoot length (6.55 and 7.99 cm), shoot diameter (1.54 and 1.68 mm), leaf number (9.00 and 9.55), and leaf area (3.91 and 5.49 cm$^2$) were found in G4 and G8, respectively. Additionally, root number was the lowest (38.33 and 38.72) in G4 and G13, respectively. Also, G13 had the shortest taproot (5.87 cm).

The analysis of chlorophylls, soluble sugars, and total phenols in the leaves and shoots of 15 *P. vera* genotypes are displayed in (Figure 7). As a result, chlorophyll *a* (Figure 7A) was significantly the maximum in G3 (34.54 µg g$^{-1}$) followed by G4 (30.44 µg g$^{-1}$), and the lowest values observed in all other genotypes, particularly in G10 (23.21 µg

g$^{-1}$). Chlorophyll *b* (Figure 7B) was the highest in G13 (24.74 µg g$^{-1}$) and the lowest was recorded in G3 (11.51 µg g$^{-1}$). In addition, the values of soluble sugar contents (Figure 7C) in shoots of *P. vera* were the greatest in G7 (510.86 µg g$^{-1}$) and G6 (459.01 µg g$^{-1}$). While, G4 and G8 were the genotypes with the least soluble sugars (245.74 and 264.26 µg g$^{-1}$, respectively). The highest total phenol content (Figure 7D) was observed in G7 (720.00 µg g$^{-1}$), however G8 exhibited the lowest total phenol (267.00 µg g$^{-1}$).

**Table 7.** Effect of *P. vera* genotypes on germination percentage and speed along with subsequent seedling growth traits

| Genotypes | Germination % | Germination speed | Shoot length (cm) | Shoot diameter (mm) | Leaf number | Leaf area ($cm^2$) | Root number | Taproot length (cm) |
|---|---|---|---|---|---|---|---|---|
| G1 | 96.67 ab | 8.90 bc | 8.53 def | 1.78 b-e | 10.65 cde | 6.04 bc | 44.22 cd | 11.32 a |
| G2 | 78.89 cde | 9.35 ab | 10.02 b-e | 2.11 a | 12.44 abc | 6.93 ab | 58.77 a-d | 10.45 ab |
| G3 | 80.00 b-e | 5.03 de | 8.25 def | 2.08 a | 10.55 cde | 5.89 bc | 48.00 bcd | 12.82 a |
| G4 | 68.89 ef | 2.46 f | 6.55 f | 1.54 e | 9.00 e | 3.91 c | 38.33 d | 11.91 a |
| G5 | 97.78 a | 11.51 a | 12.54 ab | 1.96 a-d | 11.11 a-d | 5.76 bc | 67.44 ab | 11.34 a |
| G6 | 93.33 abc | 9.79 ab | 12.19 abc | 2.06 ab | 12.67 ab | 8.70 a | 70.04 a | 14.83 a |
| G7 | 85.56 a-d | 6.98 cd | 10.17 b-e | 2.09 a | 11.77 abc | 5.88 bc | 62.22 abc | 12.58 a |
| G8 | 62.22 f | 3.82 ef | 7.99 ef | 1.68 de | 9.55 de | 5.49 bc | 63.11 abc | 13.42 a |
| G9 | 94.44 abc | 10.60 ab | 10.74 bcd | 1.99 abc | 12.33 abc | 6.25 b | 44.66 cd | 11.73 a |
| G10 | 91.11 abc | 7.23 cd | 9.71 cde | 1.85 a-d | 12.78 a | 6.42 b | 49.66 a-d | 11.10 a |
| G11 | 98.89 a | 11.38 ab | 13.92 a | 2.07 ab | 12.44 abc | 8.55 a | 67.22 ab | 10.80 a |
| G12 | 70.00 def | 4.11 ef | 7.90 ef | 1.85 a-d | 10.77 b-e | 5.75 bc | 47.77 bcd | 11.51 a |
| G13 | 78.33 cde | 5.50 de | 7.55 ef | 1.77 cde | 10.55 cde | 5.59 bc | 38.72 d | 5.87 b |
| G14 | 93.33 abc | 6.19 de | 9.91 b-e | 1.76 cde | 11.11 a-d | 6.01 bc | 55.22 a-d | 13.88 a |
| G15 | 86.67 abc | 7.18 cd | 11.33 abc | 2.08 a | 12.66 ab | 7.35 ab | 52.77 a-d | 9.80 ab |

The means in the same column with the same letter(s) had no significant differences according Duncan's Multiple Range Test (P≤0.05).

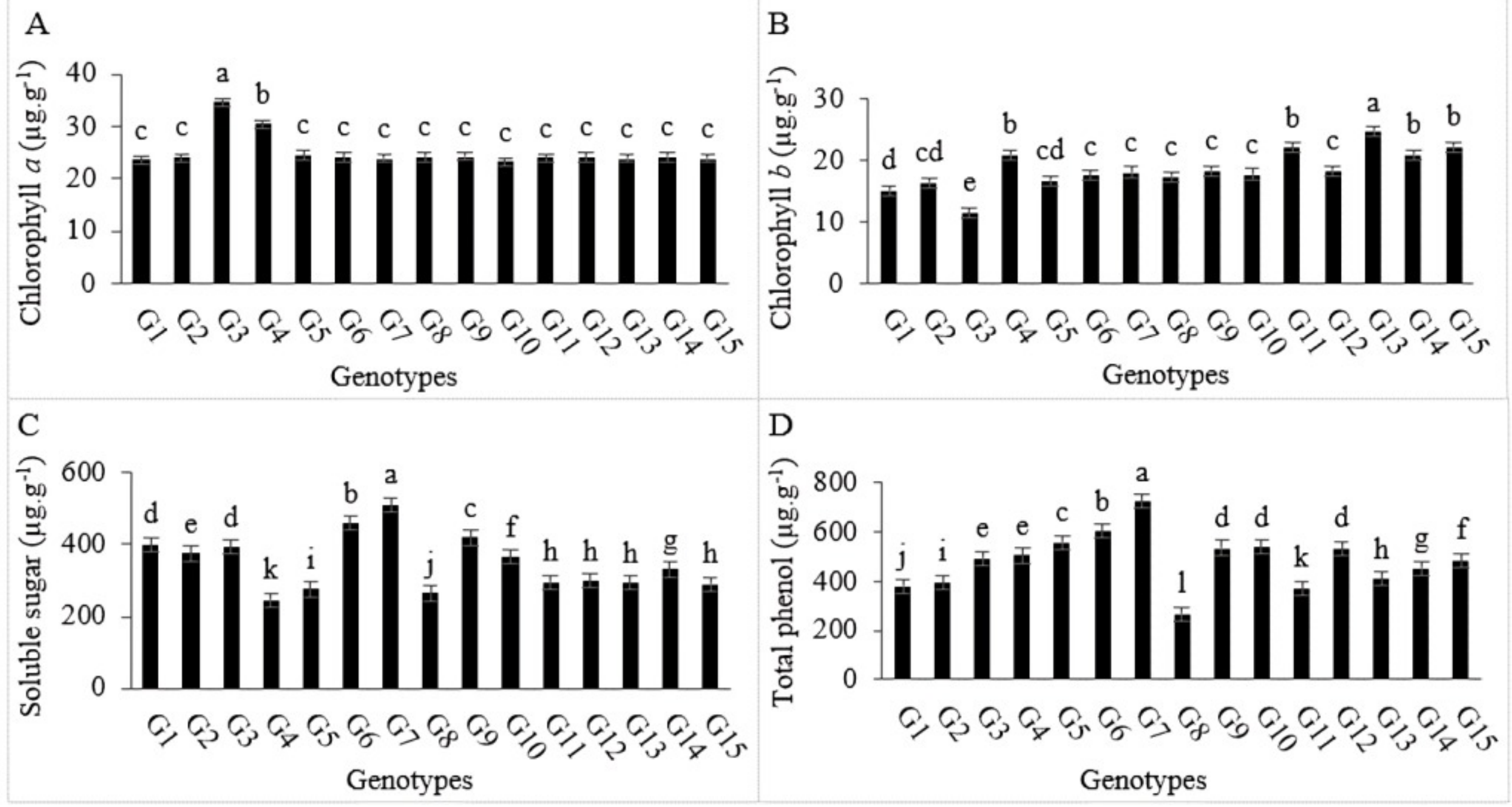

**Figure 7.** Chlorophyll a (A), chlorophyll b (B), soluble sugars (C), and total phenols (D) in seedlings of *P. vera* genotypes. The means in the same figure with the same letter denotes to no significant differences according to Duncan's Multiple Range Test (P≤0.05).

## *Interrelationships of the variances*

Principal component analysis (PCA) was conducted on all seeds and seedlings variances of the 15 *P. vera* genotypes, including seed morphological traits, seed phytochemical components, seed germination parameters, and subsequent seedling growth characteristics (Figure 8). It was found that PCA1 and PCA2 were the two components that carried 55.69% of the variables; in such a way, 33.02% for PCA1 and 22.67% for PCA2. Correspondingly, G6 and G11 at the positive side of both PCA1 and PCA2 associated with germination percentage (G%), germination speed (GS), shoot length (ShoL), leaf number (LN), leaf area (LA), shoot diameter (ShoD), root number (RN), and protein percentage in the seed kernel (Pro). Whereas, G6 and G11 negatively associated with total phenols in seed kernel (TPCse). At the same position, G6 and G7 strongly connected with seed length (SeL), total phenols content in seedling shoot (TPCsh), and soluble sugar content in seedling shoot (SSCsh). Additionally, G3 at the positive side of PCA1 assembled with seed height (SeH), seed width (SeH), seed volume (SeV), seed weight (SeWe), seed shell weight (SheWe), seed shell thickness (SheTh), seed kernel weight (KeWe), carbohydrates in the seed kernel (Carbo), soluble sugar content in the seed kernel (SSCse), and chlorophyll *a* in seedling leaves (Chl a). In contrast, G3 was in a negative connection with chlorophyll *b* (Chl b) and oil percentage in seed kernel (oil). Further, G1, G5, G13, and G14 were positively linked with antioxidants in seed kernel (DPPHse), TPCse, oil, and Chl b, but they had an inverse relationship with SeW, SeH, SeV, SeWe, SheWe, SheTh, Carbo, SSCse, and Chl a. Besides, G4 and G8 were on the negative sides of both PCA1 and PCA2, meaning that they were negatively related to the majority of variables.

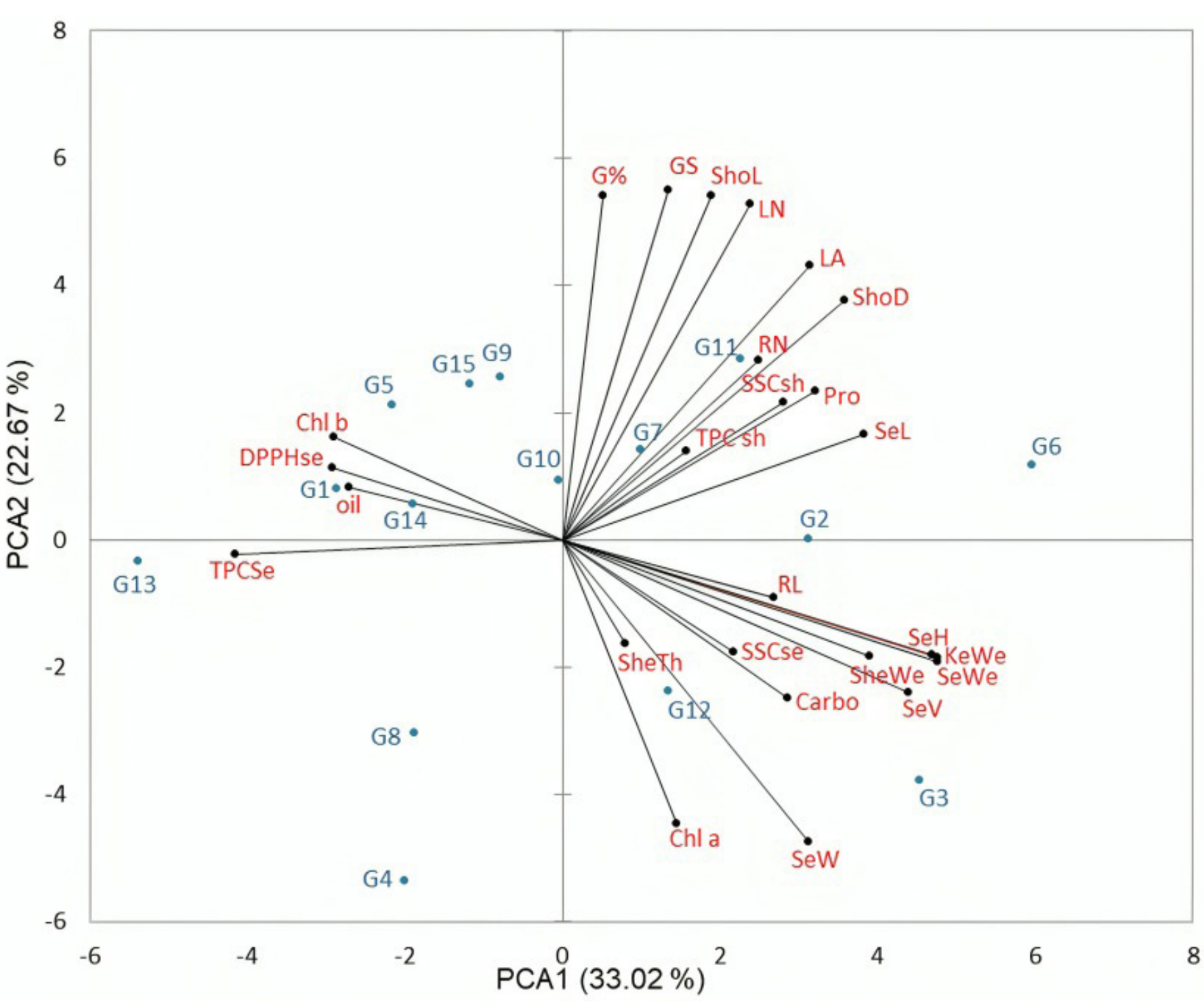

**Figure 8.** Biplot of Principal Component Analysis (PCA) of the variances. G%: germination percentage, GS: germination speed, ShoL: shoot length, ShoD: shoot diameter, LN: leaf number, LA: leaf area, RN: root number, RL: root length, Chl a: chlorophyll *a*, Chl b: chlorophyll *b*, TPCsh: total phenols content in seedling shoot, SSCsh: soluble sugars content in seedling shoot, Carbo: carbohydrates in seed kernel, oil: oil in seed kernel, Pro: protein in seed kernel, SSCse: soluble sugar content in seed kernel, TPCse: total phenol content in seed kernel, DPPHse: antioxidants in seed kernel, SeL: seed length, SeW: seed width, SeH: seed height, SeV: seed volume, SeWe: seed weight, SheWe: seed shell weight, KeWe: seed kernel weight, and SheTh: seed shell thickness.

Pearson correlation and PCA tests (Figures 8 and 9) manifested that SeL. SeW, and SeH were positively correlated with SeV, SeWe, SheWe, and KeWe (Figure 9A). Meantime, SeV and SeW were positively correlated. Further, SheWe had positive links with KeWe and SheTh. In addition, Figure (9B) indicates that Carbo and Pro had negative correlation with oil and TPCse. Pro connected with ShoD, LN, and LA positively. Oil and TPCse positively associated. TPCse correlated positively with DPPHse and negatively with LA. Chl b and RL negatively correlated. Moreover, SSCsh had a positive link with ShoD. In addition, other positive correlations; G% with (GS, ShoL, LN, and LA), GS with (ShoL, ShoD, LN, LA, and RN), ShoL with (ShoD, LN, LA, and RN), ShoD with (LN and LA), LN with (LA), and LA with (RN) were observed.

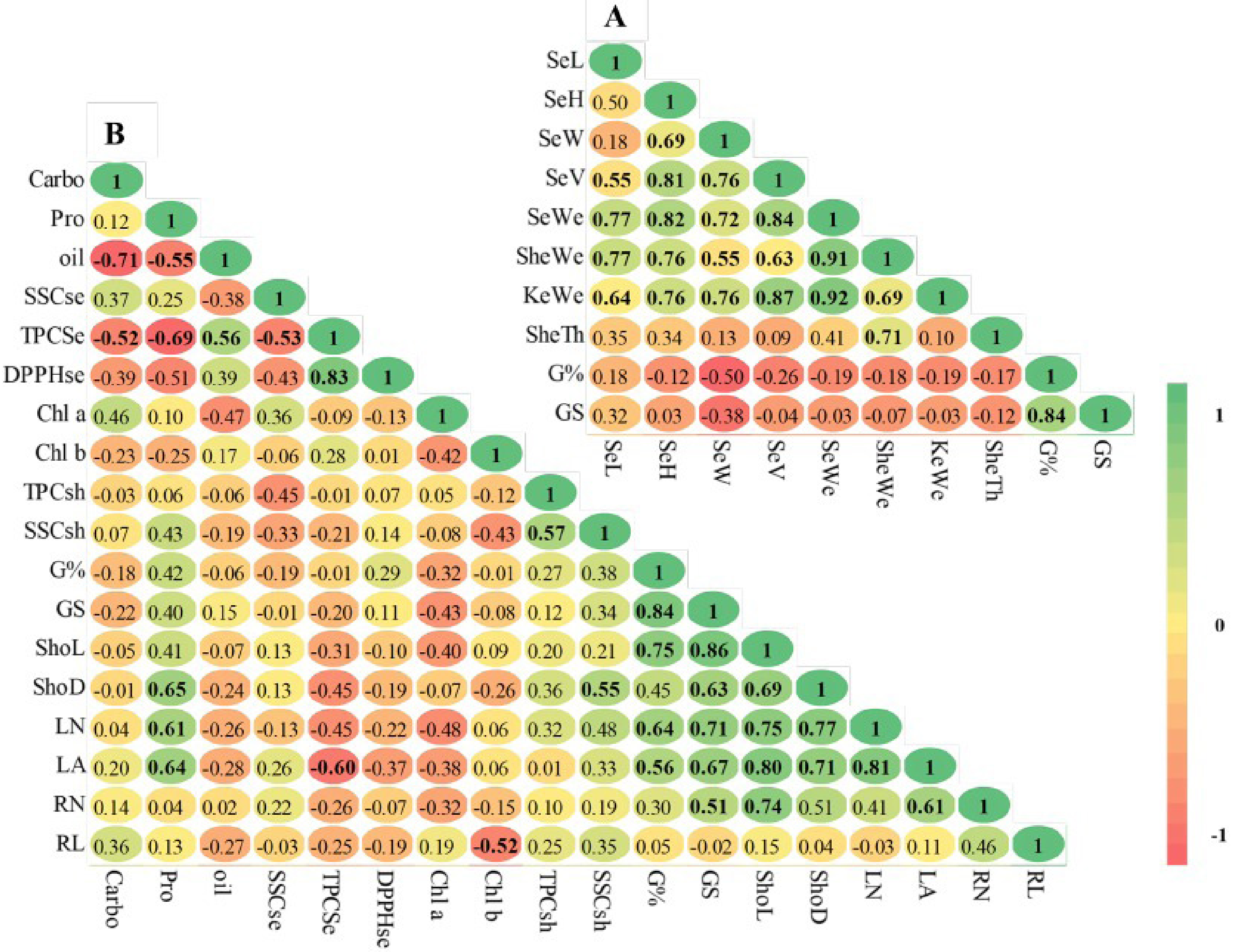

**Figure 9.** Pearson correlation test (P≤0.05) among the parameters. Correlations among seed morphological traits and germination parameters (A), and correlations among phytochemicals in seed kernel, phytochemicals in subsequent seedlings growth traits, germination parameters, and subsequent seedling growth traits (B). The bold values indicated significant correlation. G%: germination percentage, GS: germination speed, ShoL: shoot length, ShoD: shoot diameter, LN: leaf number, LA: leaf area, RN: root number, RL: root length, Chl a: chlorophyll a, Chl b: chlorophyll b, TPCsh: total phenols content in seedling shoot, SSCsh: soluble sugars content in seedling shoot, Carbo: carbohydrates in seed kernel, oil: oil in seed kernel, Pro: proteins in seed kernel, SSCse: soluble sugars content in seed kernel, TPCse: total phenols content in seed kernel, DPPHse: antioxidants in seed kernel, SeL: seed length, SeW: seed width, SeH: seed height, SeV: seed volume, SeWe: seed weight, SheWe: seed shell weight, KeWe: seed kernel weight, and SheTh: seed shell thickness.

The studied 15 *P. vera* genotypes were grouped into four clusters when hierarchical clustering analyses were conducted on the bases of seed morphological properties, seed phytochemical components, germination parameters, and seedling charac-teristics post germination (Figure 10). The first group (yellow) was G1, G14, G5, and G13. This group was recognized by high to mid GP, DPPHse, TPCse, and oil, but mid to low seed morphological properties. The second group (red) contained G9, G15, G10, and G11, and they had the best GP, LN, and Pro, but SheTh, SheWe, oil, TPCse, and Chl a were low in this group. Moreover, G4 and G8 were the third group (blue); GP, GS, and other parameters were low in this group. The fourth group (green) included G2, G12, G7, G6, and G3, in which GP and GS were high to intermediate, seed morphological properties were the best, and other traits were intermediate.

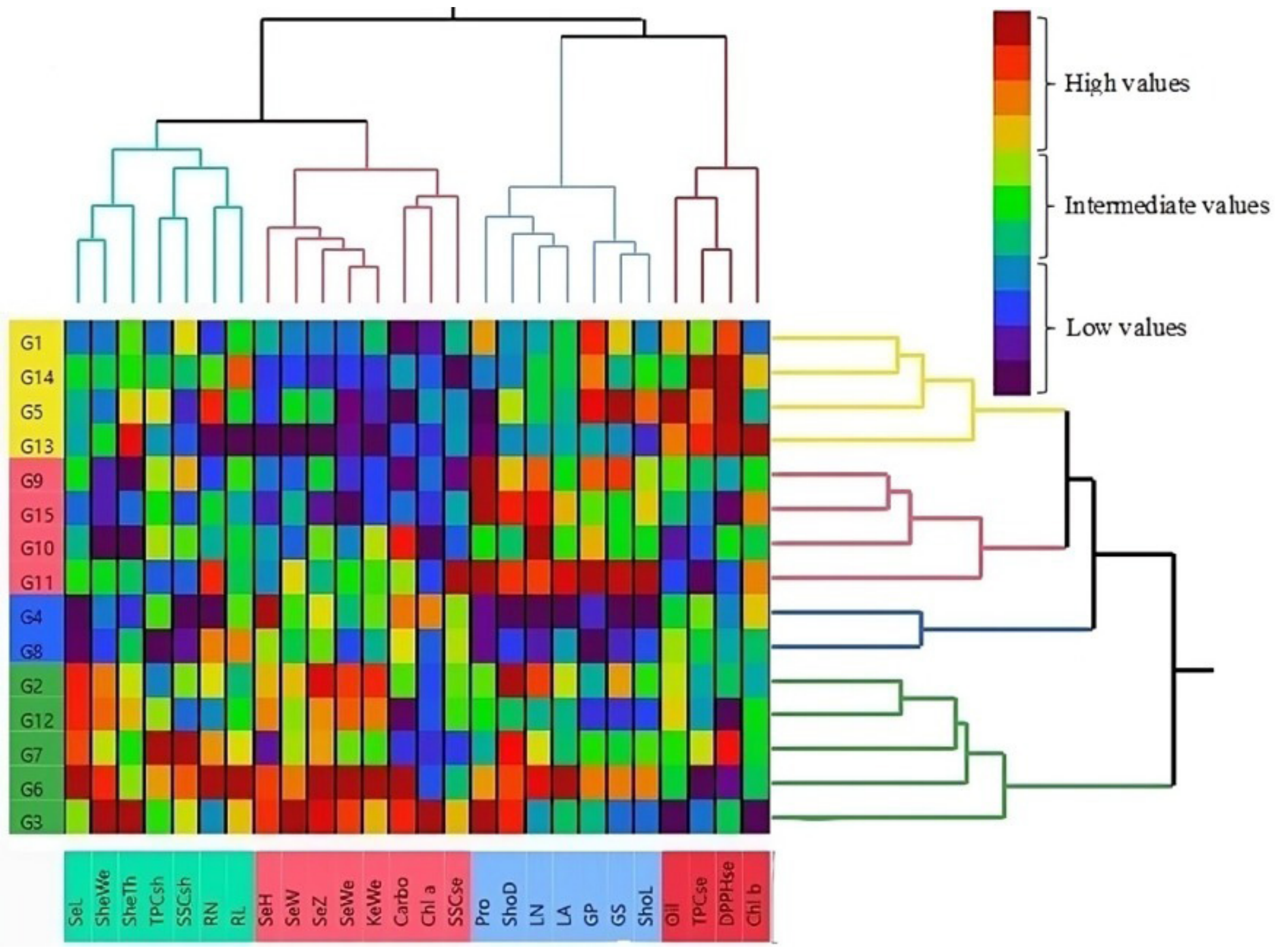

**Figure 10.** Hierarchical clustering (Ward method) based on the *P. vera* genotypes and their seed morphological traits, seed phytochemical components, germination parameters, and characteristics of seedlings subsequent germination. GP: germination percentage, GS: germination speed, ShoL: shoot length, ShoD: shoot diameter, LN: leaf number, LA: leaf area, RN: root number, RL: root length, Chl a: chlorophyll *a*, Chl b: chlorophyll *b*, TPCsh: total phenols content in seedling shoot, SSCsh: soluble sugars content in seedling shoot, Carbo: carbohydrates in seed kernel, oil: oil in seed kernel, Pro: proteins in seed kernel, SSCse: soluble sugars content in seed kernel, TPCse: total phenols content in seed kernel, DPPHse: antioxidants in seed kernel, SeL: seed length, SeW: seed width, SeH: seed height, SeZ: seed volume, SeWe: seed weight, SheWe: seed shell weight, KeWe: seed kernel weight and SheTh: seed shell thickness.

# Discussion

The purpose of the genetic diversity analysis of the 15 *P. vera* genotypes investigated in this study was to reveal the degree of variability among them and how this variability degree reflects on germination ability and other characteristics of produced seedlings post germination, along with some phytochemicals. Genetic variability can directly be uncovered through the values of Na (number of observed alleles), Ne (number of effective alleles), He (expected heterozygocity), I (Shannon's information index), and uHe (unbiased heterozygocity) (Majeed et al., 2024). In the current study, regarding the highest values of Na, Ne, He, I, and uHe, both ISSR and RAPD primers had the capacity to indicate genetic variability. UBC 834 and ISSR 9 from ISSR primers, and OPD-18, OPAW-10, and OPAV-19 from RAPD were the best primers to expose genetic diversity among the 15 *P. vera* genotypes (Table 2). However, the mean values of these indices demonstrated that RAPD primers in general were more efficient than ISSR primers in revealing genetic diversity among the 15 *P. vera* genotypes because the mean values of Na, Ne, He, I, and uHe resulted from RAPD primers were higher than those of ISSR primers. Similar results were gained when Baghizadeh et al. (2010) applied RAPD, ISSR, and SSR markers to Iranian pistachios, and they summarized that to differentiate the genotypes, RAPD markers were the most powerful ones.

The 15 *P. vera* genotypes were divided into groups and displayed in dendrograms (Figure 5), and their genetic purity was displayed in structure analysis (Figure 6) based on the highest values of K. In this regard, ISSR primers resulted in a greater dissimilarity range than RAPD ones. Four distinct groups detected as ISSR primers were used (Figure 5A) in contrast to RAPD primers that resulted in the characterization of three groups (Figure 5B). Whereas, three groups were observed in the combination of the two primers (Figure 5C). Also, four groups because of ISSR primers, three groups because of RAPD and combination of ISSR and RAPD primers were obtained in structure analysis (Figures 6B, D, and F). Correspondingly, (G1 and G13), (G2, G3, G4, G5, G6, and G7), and (G9 and G10) were always together in the same group in the dendrograms produced due to ISSR, RAPD, or combination of ISSR and RAPD primers. Depending on the value recorded between 0.2 and 0.8, structure analysis clarified that (G1, G13, and G14), (G4 and G6), and (G9 and G10) were purely existing together in the same group whenever ISSR, RAPD, or combination of ISSR and RAPD primers were applied. For demonstrating diversity between the population, the highest values of Na, Ne, He, I, and uHe (Table 3) explained that populations 4 and 3 as the result of ISSR primers, and populations 2 and 1 due to RAPD primers were the most diverse.

Comparison among the 15 *P. vera* genotypes suggested that genetic proximity between the genotypes reflected in the morphological characteristics of the seeds (Table 5). G2, G3, G6, G7, and G12 had the best seed morphological characteristics. Meanwhile, G2, G3, G6, and G7 were always together in the same group of the different dendrograms (Figure 5), and G12 was grouped with these genotypes based on the second group of combination of ISSR and RAPD primers. In the same (Table 5), it was also found that G1, G8, G13, G14, and G15 were characterized by low seed morphological traits. It was gained from the dendrogram and structure analysis that G1, G13, and G14 were always clustered, and G8 and G15 were shown within the groups of these genotypes considering dendrograms (Figures 5B and C, respectively). Besides, all seed morphological traits were statistically similar in G9 and G10, and these two genotypes were falling in the same group of all dendrograms and structure analyses. On the other hand, the Pearson correlation (Figure 9A) demonstrated that the weight of seed, shell, and kernel positively correlated with seed volume. Also, seed length, width, and height were in positive associations with seed volume. These imply that when volume increased, the accumulated matter, more space, and surface area were increased, therefore the mentioned traits increased. In the same figure, shell weight and shell thickness were positively related, and this suggests that thick shells contain high matter, thus shell weight was elevated.

It appeared that phytochemical contents in the seeds of the 15 *P. vera* genotypes had relationships with genetic proximities to some extents (Table 6). Since G2, G3, G4, and G6 contained high carbohydrates, and these genotypes were noticed together in the same group in the dendrograms of ISSR, RAPD primers and their combination. Additionally, protein was great in G11, G12, and G15, and they are located together in the same groups in all dendrograms. Furthermore, G9 and G10 had high proteins, and they were always together in the same group in all dendrograms. Contradictorily, some genotypes contained low reserved food and phytochemicals in their seeds whenever they fell together in the same group of a high specific components. For example, G9 and G10 were always together in the same group in the dendrograms and structure analysis, carbohydrates content was high in G10 but low in G9. This difference could be attributed to the influence of the phytochemicals on each other. In the current study, the correlation test elucidated that some analyzed reserved foods and phytochemicals correlated with each other positively or negatively. Protein and carbohydrates negatively associated with oil (Figure 9). Carbohydrates, proteins, and soluble sugars were negatively connected to total phenols; however, oil and total phenols were positively correlated. Antioxidants and phenols were also in a positive correlation. Hence, G3, G10, and G11 contained high carbohydrates and proteins, whereas oil was the lowest in these genotypes. Carbohydrates and proteins were the minimum in G5, in contrast it had the maximum oil. Further, percent oils of G5 and G13 were the best, and they had the maximum total phenols and antioxidants as well. Previous studies found interconnections among seed components, and the vital role in this connection has been ascribed to species (CHUNG et al., 2003; BABATUNDE et al., 2022). This may belong to that seed storage capacity is limited; when high oil accumulates in a seed, little space may remain for carbohydrates, proteins, and/or others, and vice versa (SADRAS; DENISON, 2009). On the other hand, the formation pathways of seed components are regulated differently, and the same stimulus may induce the formation of some of these components and inhibit others at the same time (YRUELA, 2015). For instance, fatty acids and phenolic compounds are concurrently produced from the shikimic acid pathway (CANDEIAS et al., 2018), and in the present study oil and total phenols positively correlated. Besides, environmental factors directly impact on the building up of the seed components (WEERASEKARA et al., 2021), the pistachio seeds used in this study came from different environments. The reason why total phenols and antioxidants positively linked likely due to phenolic compounds and neutralized free radicals through denoting hydrogen atoms and electrons, so they act as antioxidants and reduce oxidative damages (LIU et al., 2022).

Genetic closeness reflected in the germination percentages of the 15 *P. vera* genotypes investigated in this study (Table 7). The highest germination percentage was detected in G11, and it was not significantly different from G15 that they were together in the same subgroup branch according to ISSR and RAPD primer dendrograms, either individually or in combination. Whereas, G12 was in the same group of G11 and G15, but it was in an independent subgroup branch in the dendrograms. There were similarities among the majority of the genotypes in germination percentages when they were observed as being in the same subgroup branch. In this context, G4 and G8 had the lowest germination percentages; G4 was in the subgroup branch of G2 according to RAPD and the combination of ISSR and RAPD primers, which were similar in low germination percentages. Correspondingly, RAPD and the combination of ISSR and RAPD primers explained that G8 was an independent subgroup in the existing groups. Moreover, the following genotypes were similar in germination percentages and they were in the same group and subgroup branches: (G1 and G14), (G5 and G6), and (G9 and G10). However, G13 was located in the group of (G1 and G14), but it was in the subgroup branch of G12, of which the seeds germinated in a low percentage based on the dendrograms of RAPD and the combination of ISSR and RAPD primers. In connection with germination percentage and speed with seed phytochemical contents, G11, G5, G1, and G9 were respectively the top genotypes in germination percentages and speeds, and in these genotypes at least one seed component was the highest (Table 6). The maximum protein contents were measured in G1, G9, and G11, while the best oil percentages were noticed in G5 and G1, and G11 had the highest soluble sugars. These phytochemicals ordinarily provide energy and building blocks to new cells and enzymes during the germination process (AVEZUMA et al., 2023). It was worth mentioning that germination percentage positively correlated with germination speed (Figure 9B). Other seedling characteristics of the 15 *P. vera* genotypes post germination delineated that the genotypes that had high germination percentages and speeds also had the best shoot and root traits. This could be attributed to that these genotypes had better seed chemical contents that contributed to better and faster growth. Therefore, there was a positive association between protein content in the *P. vera* seeds and shoot diameter, leaf number, and leaf area in the current study. Additionally, speed of germination in those genotypes may impart the genotypes more time to do photosynthesis and increase shoot and root growth. On the other hand, germination percentage and speed did not positively or negatively relate to seed morphological characteristics (Figure 9A). In other words, gemination parameters were independent of seed morphological characteristics.

On the whole, regarding the Ward method of clustering analysis (Figure 10), the produced clusters formed as a result of the measured parameters of seed morphological traits, seed phytochemical components, germination parameters, and subsequent seedling growth characteristics were matched at the high extent with the clusters formed from genetic diversity analysis of the 15 *P. vera* genotypes, represented in the dendrograms. (G1, G13, and G14), (G9, G10, G11, and G15), and (G2, G3, G6, and G7) were in the same group in both dendrograms as a result of genetic dissimilarity and Ward method cluster analysis. Despite this, G4 and G8 appeared as an independent group. This was because the two later genotypes had nearly all lower results of the parameters.

# Conclusion

The study of genetic diversity of the 15 *P. vera* genotypes indicated that variations among the genotypes were exposed using both ISSR and RAPD primers. As a result, three constant groups of the genotypes were observed, which are (G1, G13, and G14), (G2, G3, G4, G5, G6, and G7), and (G9 and G10). However, G8, G11, G12, and G15 were in different groups depending on the type of primers. ISSR primers identified G11, G12, and G15 with (G9 and G10), but RAPD primers grouped G11, G12, and G15 with (G1, G13, and G14). Furthermore, G8 and G14 were one group according to ISSR primers, but RAPD primers grouped G8 with (G2, G3, G4, G5, G6, and G7). Genetic proximity among the genotypes seemed to play a role in similarity in seed morphological properties and seed phytochemical contents. Additionally, germination percentage and speed consequences were similar depending on the genetic closeness. High germination percentage and speed were somewhat related to the ratio of at least one high reserved food component and phytochemical content, but not to the seed morphological traits. High soluble sugars and total phenols in the seed kernel in a genotype did not associate with high soluble sugars and total phenols in seedlings of the same genotype subsequent germination. Germination percentage and speed were positively correlated; the genotypes with a high germination percentage also had the fastest germination. Subsequent germination data proved that speed in germination led to enhanced root and shoot traits in the *P. vera* genotypes. The maximum root and shoot characteristics were found in the genotypes that showed the fastest germination. Consequently, to produce high numbers of vigor seedlings of *P. vera*, G11, G5, G1, G9, G6, G14, and G10 will be the promising genotypes. Whereas, germination is a complex process, so the germination capacity in *P. vera* genotype seeds should be studied in the future from other aspects such as differences in hormonal contents and discrepant environmental factors.